\documentclass[twocolumn,nofootinbib,preprintnumbers,amsmath,amssymb]{revtex4}
\usepackage{graphicx}
\usepackage[normalem]{ulem}
\usepackage{hyperref,color}

\newcommand{\rr}{{\bf r}}
\newcommand{\jj}{{\bf j}}
\newcommand{\kk}{{\bf k}}
\newcommand{\be}{\begin{equation}}
\newcommand{\ee}{\end{equation}}
\begin{document}
\title{Comment on ``Deformations, relaxation and broken symmetries in liquids, solids and glasses:
a unified topological field theory"}

\author{Taras Bryk}
\affiliation{Institute for Condensed Matter Physics of the National Academy of Sciences of Ukraine, UA-79011 Lviv, Ukraine}
\affiliation{Institute of Applied Mathematics and Fundamental Sciences, Lviv National Polytechnic University, UA-79013 Lviv, Ukraine}

\author{Walter Schirmacher}
\affiliation{Institut f\"ur Physik, Universit\"at Mainz, D-55099 Mainz, Germany}

\author{Giancarlo Ruocco}
\affiliation{Center for Life Nano Science @Sapienza, Istituto Italiano di Tecnologia, 295 Viale Regina Elena, I-00161, Roma, Italy}
\affiliation{Dipartimento di Fisica, Universita' di Roma ``La Sapienza'', I-00185, Roma, Italy}
\begin{abstract}
We discuss 
	a field-theoretical approach
	to liquids, solids and glasses, published recently
[Phys.Rev.E {\bf105}, 034108 (2022)], which aims to describe these
materials in a common quantum formalism. We argue that such quantum
formalism is not applicable to classical liquids, and the results
presented, which rely heavily on the concept of phase relaxation borrowed
from quantum fluids, contradict the known hydrodynamic theory
of classical liquids. In particular, the authors miss the important  
particle-number conservation law and the density fluctuations as
hydrodynamic slow variable.
Instead, the authors invoke
Goldstone bosons as elementary hydrodynamic excitations.
	We point out that in a classical
liquid there are no broken continuous symmetries and consequently
no Goldstone bosons. The authors claim that the Goldstone
	bosons would be responsible for
the existence of sound in liquids, instead of resulting from combined
particle-number and momentum conservation, 
	a fact well
	documented in fluid-mechanics textbooks.
\end{abstract}

\maketitle

In a recent paper \cite{Bag22} Baggioli, Landry and Zaccone (BLZ) present
a formalism aimed to describe crystalline solids, glassy solids and
liquids in a unified way. They use concepts borrowed from a field theoretic
description of quantum liquids and solids \cite{landry},
in particular Abrikosov lattices
in type-2 superconductors and charge-density-wave systems
\cite{dela17} as well as the (``holographic'') analogy between
conformal field theory for condensed-matter 
systems near a quantum phase transition
and anti-deSitter general relativity \cite{zaanen}.

In the following we focus mainly on BLZ's description of excitations
in classical liquids and leave other aspects of their treatment aside,
such as 
attempts to map 
features of quantum
condensed-matter systems with broken continuous symmetries
onto non-affine displacements
in classical liquids and glasses.

All liquids except the helium liquids can (and must be) described
in terms of classical statistical mechanics, because their
thermal 
de-Broglie wavelength
$\lambda=h/[mk_BT]^{1/2}$ is much smaller than
the diameter of the liquid particles. Here
$h$ and $k_B$ are Planck's and Boltzmann's constants,
$m$ is the particle mass, and $T$ is the temperature. Therefore
a quantum approach, in which $\hbar=\frac{h}{2\pi}$ is set equal to unity
(as done by BLZ),
appears to be unadequate to describe non-quantum liquids, to which the
quoted literature (BLZ's References [14-24]) refers.

Any proposed hydrodynamic approach of classical liquid 
must start with naming the
conservation laws for a set of the relevant hydrodynamic variables
\cite{Lan,For,Han,Boo}. The conserved quantities are particle number, 
momentum and energy. The corresponding hydrodynamic variables,
which describe the collective modes,
are the number density $n(\rr,t)$, the mass-current density $\jj(\rr,t)$ and
the energy density $e(\rr,t)$. 
It is important to notice,
that the damping of each collective
hydrodynamic mode is proportional to $k^2$ \cite{Lan,For,Han,Boo} with $k$ being the wave number, 
i.e. their lifetime tends to infinity in the long-wavelength (continuum) limit as the consequence 
of local conservation laws.  

However, in the treatment of BLZ
\cite{Bag22} the density does not appear as a relevant field, 
nor does the corresponding continuity equation. Instead of the
hydrodynamic longitudinal
sound mode, which arises from particle number and momentum
conservation, BLZ invoke Goldstone bosons as elementary hydrodynamic
excitations,
which usually correspond to phase fields
arising from a broken gauge symmetry \cite{dela17}.
This results in a longitudinal hydrodynamic matrix (Equation (38) of BLZ), which
contradicts the known hydrodynamic modes in fluids
\cite{Lan,For,Han,Boo}.

Here, we review the correct form
of the hydrodynamic matrix for longitudinal dynamics, the eigenmodes 
of which are the hydrodynamic modes in the longitudinal channel. 
For describing the longitudinal excitations of classical liquids
one may use the
three orthogonal
dynamic variables
$n(\kk,t), j^{\scriptscriptstyle
L}(\kk,t), h(\kk,t)$,
where the first two
are the Fourier components of the number 
density and longitudinal mass-current density fluctuations.
$h(\kk,t)$ denotes fluctuations of the heat density, which is
the energy density, orthogonalized to the number density \cite{Bry10}
\begin{equation} \label{hkt}
h(k,t)=e(k,t)-\frac{\langle e_{-k}n_{k}\rangle}{\langle n_{-k}n_{k}\rangle}n(k,t),
\end{equation}
where the brackets denote a statistical average. Using 
this set of orthogonal hydrodynamic variables,
one 
obtains the following hydrodynamic $3\times 3$ matrix
\begin{equation}
\begin{array}{l}
	{\bf T}^{(hyd)}
	(k)= \left(
\begin{array}{ccc}
      0    & -ikc_T               &    0                 \\
    -ikc_T &  D_Lk^2              & -ikc_T\sqrt{\gamma-1}\\
      0    & -ikc_T\sqrt{\gamma-1}&  \gamma D_Tk^2       \\
\end{array}
\right)\ , \label{Th}
\end{array}
\end{equation}
where $c_T$ is the isothermal speed of sound, $\gamma=C_P/C_V$ is the ratio of the specific 
heats. $D_L$ is the
longitudinal kinematic viscosity and $D_T$ the thermal diffusivity (diffusivity of the local temperature).
One can see that when $\gamma=1$ (no coupling between the thermal and viscous processes) the 
eigenmodes of ${\bf T}^{(hyd)}(k)$ can be estimated immediately:  in this
case the hydrodynamic matrix has
one purely real eigenvalue
\begin{equation}\label{z_th}
z_{th}(k)=D_Tk^2\,,
\end{equation}
which corresponds to a thermal relaxation mode, and a pair of complex-conjugated eigenvalues
$$
z_{\pm}(k)=\frac{D_L}{2}k^2\pm i c_Tk,\qquad  (for~\gamma=1),
$$
which are the sound modes, which are
decoupled from the thermal fluctuations with linear dispersion and
isothermal speed. In the general case of $\gamma \geq 1$ the standard sound modes with adiabatic
speed of sound $c_s=c_T\sqrt{\gamma}$ and correct hydrodynamic damping 
\begin{equation}\label{z_s}
z_{\pm}(k)=\frac{D_L+(\gamma-1)D_T}{2}k^2\pm i c_sk
\end{equation}
are obtained from the hydrodynamic matrix (\ref{Th}). The origin of the long-wavelength sound modes 
is now clearly seen  even in the
particular case of $\gamma=1$: they come from the coupling of density and mass-current 
fluctuations. At variance, in \cite{Bag22} BLZ do not account for the density 
fluctuations
as the hydrodynamic variable (the set of dynamic variables in their Eq. (14), i.e. they ignore the continuity equation 
\be\label{cont}
i\omega \rho(\kk,\omega)+kj^{L}(\kk,\omega)=0\, ,
\ee
which is fundamental for all liquids \cite{Lan,For,Han,Boo}.
Recently some of us \cite{Bry21} have demonstrated that one obtains
results, which contradict the known dynamic properties of liquids,
if the continuity equation is not taken into account.

As a consequence of the absence of the continuity equation (\ref{cont})
in their treatment,
BLZ claim that their longitudinal sound would come
``from the mixing of energy fluctuations and longitudinal momentum
fluctuations'' (quoting BLZ \cite{Bag22}),
at variance with textbook knowledge \cite{Lan,For,Han,Boo}.

Further, BLZ do not obtain the thermal relaxation mode 
(\ref{z_th}) among the eigenmodes of their $3\times 3$ "hydrodynamic matrix" (their Eq. (38) and
Fig.9), 
that means, that they do not recover the standard Rayleigh-Brillouin three-peak shape for the 
dynamic structure factors $S(k,\omega)$ (see Fig. 4.2 in Forster's book
\cite{For})
with the famous Landau-Placzek ratio \cite{Lan,For,Han,Boo}. 
Their single relaxation mode behaves (in their Fig.9) as typical non-hydrodynamic relaxation,
which has a finite lifetime at large length scales,
and cannot contribute to 
$S(k\to 0,\omega)$ in the hydrodynamic $k\rightarrow$ 0 limit. 

We further mention that BLZ 
present a viscoelastic equation 
at variance with the
correct one \cite{maxwell,dyre06}
\be
\frac{d}{dt}\gamma(t)=\frac{1}{G}\frac{d}{dt}\sigma(t)+\frac{1}{\eta}\sigma(t)
=\frac{1}{G}\bigg(\frac{d}{dt}+\frac{1}{\tau}\bigg)\sigma(t)
\ee
where $\gamma(t)$ is the strain rate, $\sigma(t)$ the stress,
$G$ the shear modulus, $\eta$ the shear viscosity and
$\tau=\eta/G$ Maxwell's relaxation time. BLZ incorrectly
call $\tau$
``single-particle relaxation time'' 
(in caption to their Fig.5),
wheras it describes the relaxation of the {\it collective macroscopic}
shear stress $\sigma(t)$ \cite{Bry18}.

In the end we would like to emphasize that in a classical
liquid there is no phase relaxation as in condensed quantum
systems \cite{dela17}. BLZ assign phase relaxation, i.e. the relaxational
behavior of the postulated Goldstone bosons, an important role in their
treatment.

We mention here, that an
 account for stress tensor fluctuations beyond hydrodynamics
 \cite{Bry10} 
provides a clear 
picture of $k$-dependent Maxwell relaxation as well as of the origin of the positive sound 
dispersion, instead invoking BLZ's phase relaxation. 
This treatment has been further shown to fulfil a higher number
of exact sum rules than the pure hydrodynamics \cite{Bry10}.

{\it Acknowledgments}
TB was supported by NRFU grant agreement 2020.02/0115 .

\end{document}